\documentstyle[aps, prl, twocolumn, floats, epsfig]{revtex}

\newcommand{\Sign}{\mbox{Sign}}
\begin{document} \draft 
\preprint{DUKE-TH-99-183, MIT-CTP 2821/99}
\title{Meron-Cluster Solution of Fermion Sign Problems}
\author{Shailesh Chandrasekharan$^\dagger$ and Uwe-Jens Wiese$^\ddagger$}
\address{$^\dagger$ Department of Physics, Box 90305, Duke University,
Durham, NC 27708, U.S.A.}
\address{$^\ddagger$ Center for Theoretical Physics,
Massachusetts Institute of Technology,
Cambridge, MA 02139, U.S.A.}
\date{February 10, 1999}
\twocolumn[\hsize\textwidth\columnwidth\hsize\csname @twocolumnfalse\endcsname
\maketitle

\begin{abstract}
We present a general strategy to solve the notorious fermion sign problem using
cluster algorithms. The method applies to various systems in the Hubbard model 
family as well as to relativistic fermions. Here it is illustrated for 
non-relativistic lattice fermions. A configuration of fermion world-lines is 
decomposed into clusters that contribute independently to the fermion 
permutation sign. A cluster whose flip changes the sign is referred to as a 
meron. Configurations containing meron-clusters contribute $0$ to the path
integral, while all other configurations contribute $1$. The cluster 
representation describes the partition function as a gas of clusters in the 
zero-meron sector.
\end{abstract} 
]%% end of \twocolumn
%
 
%\narrowtext 

The numerical simulation of fermions is a notorious problem that hinders
progress in understanding high-temperature superconductivity \cite{Whi89}, QCD 
at non-zero chemical potential \cite{Bar98} and many other important problems 
in physics. One of the main problems originates from the minus-signs associated
with Fermi statistics which prevent us from interpreting the Boltzmann factor 
in a fermionic path integral as a positive probability. When the sign of the 
Boltzmann factor is incorporated in measured observables, the fluctuations in 
the sign give rise to dramatic cancellations. Especially for large systems at 
low temperatures this leads to relative statistical errors that are 
exponentially large in both the volume and the inverse temperature. This makes 
it impossible in practice to study such systems with standard numerical 
methods. Here, for the first time, we completely eliminate a severe sign 
problem in the simulation of a non-relativistic system of interacting lattice 
fermions using a cluster algorithm. The solution of the problem proceeds in two
steps. The idea of the first step is to use cluster algorithm techniques to 
reduce the problem of canceling many contributions $\pm 1$ to the problem of 
averaging over non-negative contributions $0$ and $1$. This step solves one 
half of the sign problem as discussed below. In large volumes and at small 
temperatures one still generates vanishing contributions to the average sign 
most of the time and very rarely one encounters a contribution $1$. In order to
solve the other half of the problem a second step is necessary which guarantees
that contributions $0$ and $1$ are generated with similar probabilities. The 
idea behind the second step is to include a Metropolis decision in the process 
of cluster decomposition. The two basic ideas behind our algorithm are general 
and apply to a variety of systems. In this paper, we illustrate our method for 
a simple model which serves as a testing ground for the new ideas. 

Let us consider a fermionic path integral $Z_f = \sum_n \Sign[n] \exp(-S[n])$ 
over configurations $n$ with a Boltzmann weight of $\Sign[n] = \pm 1$ and 
magnitude $\exp(-S[n])$. Here $S[n]$ is the action of a corresponding bosonic 
model with partition function $Z_b = \sum_n \exp(-S[n])$. A fermionic 
observable $O[n]$ is obtained in a simulation of the bosonic ensemble as
\begin{equation}
\langle O \rangle_f = \frac{1}{Z_f} \sum_n O[n] \Sign[n] \exp(-S[n]) =
\frac{\langle O \ \Sign \rangle}{\langle \Sign \rangle}.
\end{equation}
The average sign in the simulated bosonic ensemble is
\begin{equation}
\langle \Sign \rangle = \frac{Z_f}{Z_b} = \exp(- \beta V \Delta f).
\end{equation}
The last equality (valid for large $\beta V$) points to the heart of the sign 
problem. The expectation value of the sign is exponentially small in both the 
volume $V$ and the inverse temperature $\beta$ because the difference between 
the free energy densities $\Delta f = f_f - f_b$ of the fermionic and bosonic 
systems is necessarily positive. Even in an ideal simulation of the bosonic 
ensemble which generates $N$ completely uncorrelated configurations, the 
relative statistical error of the sign (again for large $\beta V$) is
\begin{equation}
\frac{\Delta \Sign}{\langle \Sign \rangle} = 
\frac{\sqrt{\langle \Sign^2 \rangle - \langle \Sign \rangle^2}}
{\sqrt{N} {\langle \Sign \rangle}} = \frac{\exp(\beta V \Delta f)}{\sqrt{N}}.
\end{equation}
Here we have used $\Sign^2 = 1$. To determine the average sign with sufficient 
accuracy one needs to generate on the order of $N = \exp(2 \beta V \Delta f)$ 
configurations. For large volumes and small temperatures this is impossible in
practice. 

It is possible to solve one half of the problem if one can match all
contributions $-1$ with $1$ to give $0$, such that only a few unmatched 
contributions $1$ remain. Then effectively $\Sign = 0,1$ and hence 
$\Sign^2 = \Sign$. This reduces the relative error to
\begin{equation}
\frac{\Delta \Sign}{\langle \Sign \rangle} = 
\frac{\sqrt{\langle \Sign \rangle - \langle \Sign \rangle^2}}
{\sqrt{N'} {\langle \Sign \rangle}} = 
\frac{\exp(\beta V \Delta f/2)}{\sqrt{N'}}.
\end{equation}
One gains an exponential factor in statistics, but one still needs to generate 
$N' = \sqrt{N} = \exp(\beta V \Delta f)$ independent configurations in order to
accurately determine the average sign \cite{Evertz}. This is because one 
generates exponentially many vanishing contributions before one encounters a 
contribution $1$. 

In several cases cluster algorithms provide an explicit matching of 
contributions $-1$ and $1$ using an improved estimator. Cluster algorithms are 
a very efficient tool to simulate quantum spin systems 
\cite{Wie92,Eve93,Wie94}. In particular, the method can be implemented directly
in the Euclidean time continuum \cite{Bea96}. The basic idea behind these 
algorithms is to decompose a configuration into $N_C$ clusters of spins which 
can be flipped independently. Averaging analytically over the $2^{N_C}$ 
configurations generated by the cluster flips, one can construct improved 
estimators for various physical quantities. As we will show, using an improved 
estimator for the fermion sign, cluster algorithms can solve the sign problem 
if the clusters contribute independently to the sign and a reference cluster 
orientation with a positive weight always exists. This means that the flip of 
any given cluster either changes the sign or not, independent of the 
orientation of all the other clusters. A cluster algorithm for lattice fermions
was first presented in \cite{Wie93} with the hope of finding such an improved 
estimator. Unfortunately, in that algorithm the clusters do not affect the sign
independent of one another. Still, cluster algorithms have been used for 
fermion models \cite{Amm98}. For systems with no severe sign problem these 
algorithms work much better than standard numerical methods, but they do not 
solve the fermion sign problem. 

A solution to a sign problem using cluster algorithms was first found in a 
bosonic model with a complex action --- the 2-d $O(3)$ model at vacuum angle 
$\theta = \pi$ \cite{Bie95}. The cluster independence was achieved by 
constructing a non-standard action. In that model clusters whose flip changes 
the sign are half-instantons which are usually referred to as merons. In this 
paper we extend the meron-concept to fermionic models by demanding cluster
independence. For non-relativistic spinless fermions hopping on a 
$d$-dimensional cubic lattice of size $V=L^d$ with periodic boundary 
conditions, this leads us to the Hamiltonian \cite{Cha99}
\begin{equation}
H = \sum_{x,i} [- \frac{t}{2}(c_x^+ c_{x+\hat i} + c_{x+\hat i}^+ c_x) + 
U (n_x - \frac{1}{2})(n_{x+\hat i} - \frac{1}{2})],
\end{equation}
with $U \ge t > 0$. Here $\hat i$ is a unit vector in the $i$-direction, 
$c_x^+$ and $c_x$ are fermion creation and annihilation operators obeying the 
standard anticommutation relations and $n_x = c_x^+ c_x$ is the occupation 
number of the lattice site $x$. Since $U > 0$, two fermions or two holes on 
neighboring lattice sites repel each other, while a fermion and a hole attract 
one another. This is a simple example of a fermionic model for which the sign 
problem can be solved completely using a meron-cluster algorithm.

Let us now discuss our model and algorithm in more detail. Following 
\cite{Wie93} we introduce a space-time lattice with $2dM$ time-slices and 
spacing $\varepsilon = \beta/M$ in the Euclidean time direction and we insert 
complete sets of occupation number $n(x,t) = 0,1$ eigenstates at each 
time-slice to express the partition function as a path integral. The magnitude 
$\exp(-S[n])$ of the Boltzmann factor is a product of four-site interactions 
associated with space-time plaquette configurations 
$[n(x,t),n(x+\hat i,t),n(x,t+1),n(x+\hat i,t+1)]$. The sign factor 
$\Sign[n] = \pm 1$ has a topological meaning. The occupied sites form fermion 
world-lines which are closed in Euclidean time. Particles may be exchanged 
during their Euclidean time evolution and the fermion world-lines define a 
permutation of particles. According to the Pauli principle, $\Sign[n]$ is just 
the sign of that permutation. In the following we restrict ourselves to 
$U = t$. Then the bosonic system without the sign factor is the 
antiferromagnetic spin 1/2 quantum Heisenberg model, and $n(x,t) = 0$ and $1$ 
correspond to spin $- 1/2$ and $1/2$, respectively. The staggered occupation 
(the analog of the staggered magnetization)  
$O[n] = \epsilon \sum_{x,t} (-1)^{x_1+x_2+...+x_d} (n(x,t) - \frac{1}{2})$, 
and the corresponding susceptibility 
$\chi = \langle O^2 \Sign \rangle/\beta V \langle \Sign \rangle$ are important 
observables.

The algorithm decomposes a configuration into  closed loops of lattice points 
which may be flipped independently. When a loop is flipped, the occupation 
numbers of all points on the loop are changed from $0$ to $1$ and vice versa. 
Each lattice point participates in two space-time plaquette interactions 
$[n(x,t),n(x+\hat i,t),n(x,t+1),n(x+\hat i,t+1)]$. On each interaction 
plaquette the lattice points are connected in pairs and a sequence of connected
points defines a loop-cluster. For space-time plaquette configurations 
$[0,0,0,0]$ and $[1,1,1,1]$ the lattice points are connected with their 
time-like neighbors, for configurations $[0,1,1,0]$ and $[1,0,0,1]$ they are 
connected with their space-like neighbors and for configurations $[0,1,0,1]$ 
and $[1,0,1,0]$ they are connected with their time-like neighbors with 
probability $p = 2/(1 + \exp(\epsilon U))$ and with their space-like neighbors 
with probability $1-p$. After identifying the clusters, they are flipped 
independently with probability $1/2$. 

A remarkable property of the cluster rules is that 
$\Sign[n] = \prod_{i=1}^{N_C} \Sign[C_i]$, where $C_i, i = 1,...,N_C$ denotes 
the oriented clusters in a configuration. By properly flipping the clusters, 
one can reach a reference configuration (the first configuration in figure 1)
in which all even lattice sites are occupied and all odd sites are empty. In 
the reference orientation $\Sign[C_i] = 1$. When the cluster is flipped, 
$\Sign[C_i] = 1$ if $N_w + N_h/2$ is odd and $- 1$ otherwise. Here 
$N_w$ is the temporal cluster winding number and $N_h$ is the number of times
the cluster hops to a neighboring lattice point. This relation follows directly
from the fermionic anticommutation relations. Following \cite{Bie95}, we 
refer to clusters whose flip changes the sign as merons. The flip of a 
meron-cluster permutes the fermions and changes the topology of the fermion 
world-lines. Since flipping all clusters does not change the fermion sign, the 
number of meron-clusters is always even. Two fermion configurations together 
with a meron-cluster are shown in figure 1.
\begin{figure}[ht]
\begin{center}
\vbox{

\hspace{1.2cm}${\rm Sign[n]}=1$

\vspace{-0.3cm}
\epsfig{file=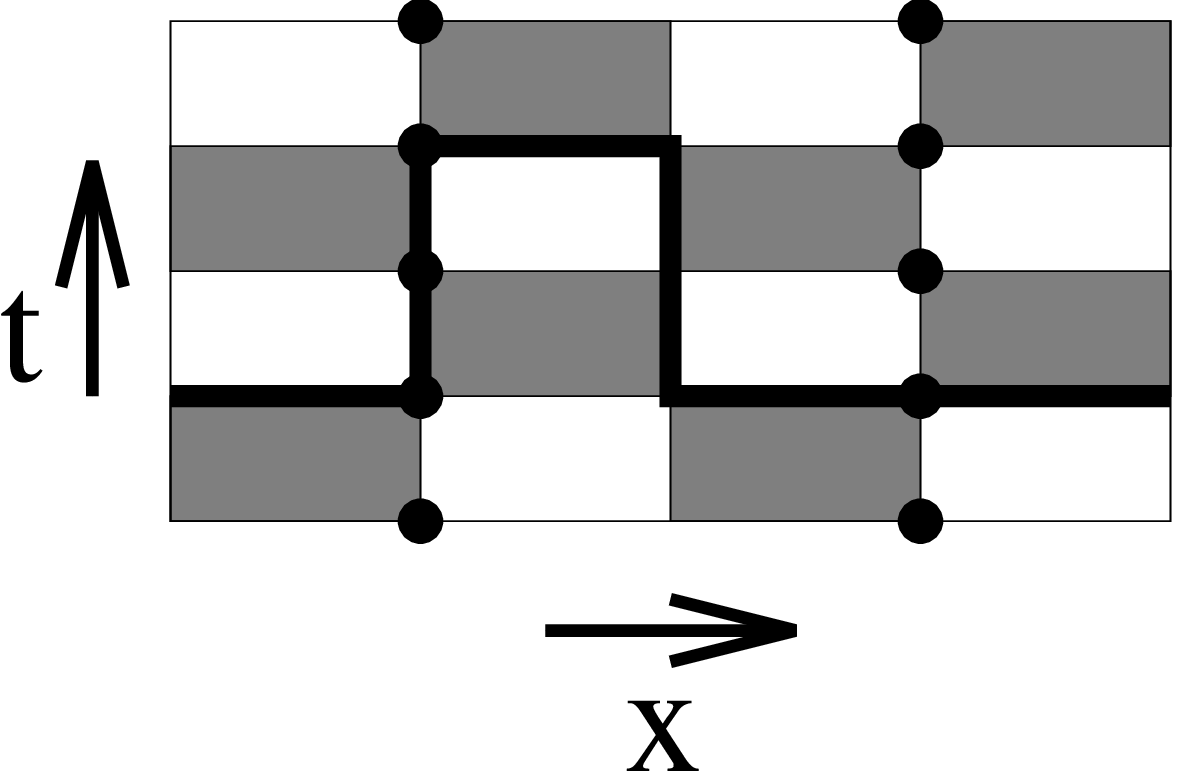,
width=4.5cm,angle=270,
bbllx=0,bblly=0,bburx=225,bbury=337}

\vspace{0.2cm}
\hspace{1.2cm}${\rm Sign[n]}=-1$

\vspace{-0.3cm}
\hspace{1.8cm}\epsfig{file=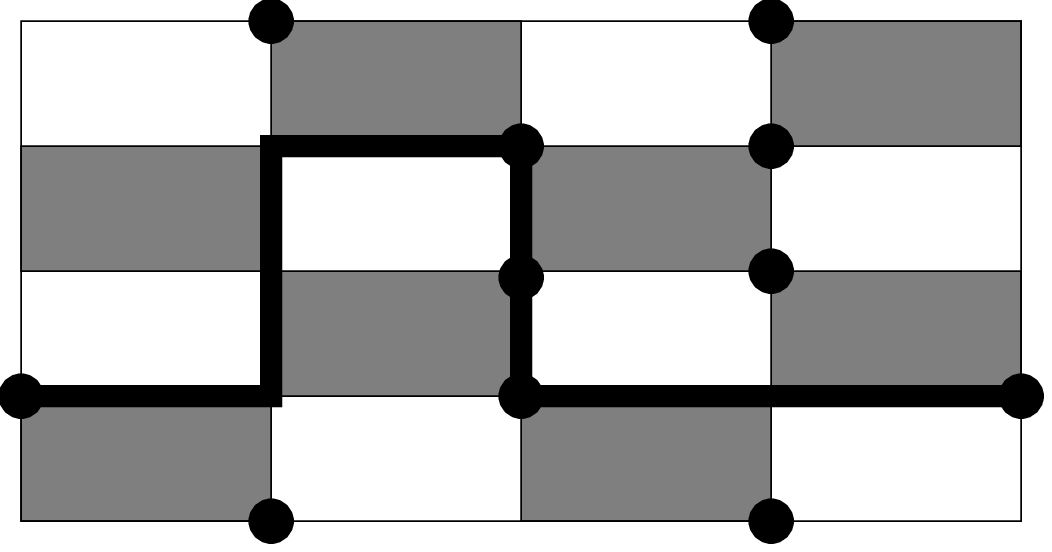,
width=4.5cm,angle=270,
bbllx=0,bblly=0,bburx=225,bbury=337}
}  
\end{center}
\vspace{-1cm}
\caption{\it  Two configurations of fermion occupation numbers in $(1+1)$ 
dimensions. The shaded plaquettes carry the interaction. The dots represent 
occupied sites. In the second configuration two fermions interchange their 
positions. Flipping a meron-cluster (represented by the fat line) changes one 
configuration into the other and changes the fermion sign. The other clusters
in the configurations are not shown.
}
\end{figure}

The improved estimator for $\langle \Sign\rangle$ is the average over the 
$2^{N_C}$ configurations obtained from independently flipping the $N_C$ 
clusters in all possible ways. For configurations that contain merons the 
average sign is zero because flipping a single meron leads to a cancellation of
signs $\pm 1$. Only the configurations without merons contribute to 
$\langle\Sign\rangle$ and their contribution is always $1$. This solves one 
half of the sign problem as discussed before. 

Let us now consider an improved estimator for $\langle O^2\Sign\rangle$ which 
is needed to determine the susceptibility $\chi$. The staggered occupation, 
$O[n] = \sum_C O_C$, is a sum of staggered occupations of the clusters, 
$O_C = \epsilon \sum_{(x,t) \in C} 
(-1)^{x_1+x_2+...+x_d} (n(x,t) - \frac{1}{2})$. When
a cluster is flipped, its staggered occupation changes sign. In a configuration
without merons, where $\Sign[n] = 1$ for all relative cluster flips, the 
average of $O[n]^2 \Sign[n]$ over all $2^{N_C}$ configurations is 
$\sum_C |O_C|^2$. For configurations with two merons the average is 
$2 |O_{C_1}||O_{C_2}|$ where $C_1$ and $C_2$ are the two meron-clusters. 
Configurations with more than two merons do not contribute to 
$\langle O^2 \Sign \rangle$. Thus, the improved estimator for the 
susceptibility is given by
\begin{equation}
\label{chi}
\chi = \frac{\langle \sum_C |O_C|^2 \delta_{N,0} + 2 |O_{C_1}||O_{C_2}| 
\delta_{N,2} \rangle}{V \beta \langle \delta_{N,0} \rangle},
\end{equation}
where $N$ is the number of meron-clusters in a configuration. Hence, to 
determine $\chi$ one must sample the zero- and two-meron sectors only.

The probability to find a configuration without merons is exponentially small 
in the space-time volume since it is equal to $\langle \Sign \rangle$. Thus, 
although we have increased the statistics tremendously with the improved 
estimators, without a second step one would still need an exponentially large 
statistics to accurately determine $\chi$. One goal of the second step is to 
eliminate all configurations with more than two merons. This enhances both the
numerator and the denominator in eq.(\ref{chi}) by an exponentially large
factor, but leaves their ratio unchanged. We start with an initial 
configuration with zero or two merons. For example, a completely occupied 
configuration has no merons. We then visit all plaquette interactions one after
the other and choose new cluster connections between the four sites according 
to the cluster rules. If the new connection increases the number of merons 
beyond two, it is not accepted and the old connection is kept for that 
plaquette. To decide if the meron number changes, one needs to examine the
clusters affected by the new connection. Although this requires a computational
effort proportional to the cluster size (and hence to the physical correlation
length) this is no problem, because one gains a factor that is exponentially
large in the volume. The above procedure obeys detailed balance because 
configurations with more than two merons do not contribute to the observables 
we consider. Also, one can show that the algorithm is still ergodic. The simple
reject step eliminates almost all configurations with weight $0$ and is the 
essential step to solve the other half of the fermion sign problem.

Since for large space-time volumes the two-meron sector is much larger than the
zero-meron sector, without further improvements one would still need statistics
quadratic (but no longer exponential) in the space-time volume to accurately 
measure $\chi$. The remaining problem can be solved with a re-weighting 
technique similar to the one used in \cite{Bie95}. To enhance the zero-meron 
configurations in a controlled way, we introduce a trial probability $p_t(N)$ 
for each $N$-meron sector. We set $p_t(N)$ for $N>2$ to infinity and use it in 
a Metropolis accept-reject step for the newly proposed cluster connection on a 
specific plaquette interaction. A new connection that changes the meron number 
from $N$ to $N'$ is accepted with probability 
$p = \mbox{min}[1,p_t(N)/p_t(N')]$. In particular, configurations with $N'>2$ 
are never generated because then $p_t(N') = \infty$ and $p=0$. After visiting 
all plaquette interactions, each cluster is flipped with probability $1/2$ 
which completes one update sweep. After re-weighting, the zero- and two-meron 
configurations appear with similar probabilities. This completes the second 
step in our solution of the fermion sign problem. The re-weighting of the 
zero- and two-meron configurations is taken into account in the final 
expression for the susceptibility as
\begin{equation}
\chi = \frac{\langle \sum_C |O_C|^2 \delta_{N,0} \ p_t(0) + 
2 |O_{C_1}||O_{C_2}| \delta_{N,2} \ p_t(2) \rangle}
{V \beta \langle \delta_{N,0} \ p_t(0) \rangle}.
\end{equation}

\begin{table}
\begin{center}
\begin{tabular}{|c|c|c|c|c|c|}
\hline
$L$ & $\beta U$ & $4M$ & $\langle \Sign \rangle$ ($\mbox{A}_1$) & 
$\chi$ ($\mbox{A}_1$) & $\chi$ ($\mbox{A}_2$) \\
\hline
\hline
  6 & 1.0 &  64 &   0.696(1) & 13.44(2) &  13.43(2) \\
\hline
  8 & 1.0 &  64 &   0.536(3) & 13.53(3) &  13.52(3) \\
\hline
  8 & 2.0 & 128 &  0.0164(5) &   199(3) &    203(2) \\
\hline
  8 & 4.0 & 256 & 0.00051(7) & 690(100) &    729(9) \\
\hline
 12 & 8.0 & 512 &    ---     &   ---    & 3090(130) \\
\hline
\end{tabular}
\end{center}
\caption{\it Numerical results for $\langle \Sign \rangle$ and $\chi$ obtained 
with algorithm $\mbox{A}_1$ and $\chi$ obtained with algorithm $\mbox{A}_2$.}
\end{table}

We have implemented the meron cluster algorithm in (2+1) dimensions and
have tested it using exact diagonalization results on small lattices.
Table 1 contains a comparison of results obtained with two algorithms using the
same number of sweeps in both cases. The first algorithm ($\mbox{A}_1$) has the
improved estimators and solves one half of the sign problem. The second 
algorithm ($\mbox{A}_2$) has both the improved estimators and the additional 
Metropolis step and also solves the other half of the problem. The algorithm
$\mbox{A}_2$ is clearly superior once the average sign becomes small. In 
particular, we have applied $\mbox{A}_2$ to systems of size $V = 12^2$ at a low
temperature $\beta U = 8$. This is far beyond reach of standard fermion 
algorithms and even of the algorithm $\mbox{A}_1$. It should be noted that our 
model has a very severe sign problem which persists after integrating out the 
fermions even at half-filling.

Cluster representations in general and the meron-concept in particular are more
than mere algorithmic tools. In fact, we have shown that the fermionic 
partition function can be expressed as a classical statistical mechanics system
of clusters. The cluster formulation is a novel type of bosonization which 
works in any dimension. In this formulation the Pauli principle manifests 
itself by the vanishing Boltzmann weight of a configuration containing 
meron-clusters. If we ignore the fermion permutation sign, the theory describes
a gas of merons and non-merons with a large configuration space. Including the 
sign factor forces even numbers of merons to be bound into non-merons. As a 
consequence, in agreement with the Pauli principle, the configuration space is 
very restricted. The merons allow us to simulate fermions with local bosonic 
variables. This is much more efficient than integrating out the fermions, which
leads to non-local bosonic effective actions.

While the details of our algorithm are specific to the fermion model we have 
considered, the two basic ideas behind it are general and apply to a variety of
models. They lead to a complete solution of the fermion sign problem for models
of relativistic staggered fermions \cite{Cha99} as well as for non-relativistic
fermions with spin. In applications of the meron-cluster algorithm to systems 
in the Hubbard model family we have so far not found high-temperature 
superconductivity. Meron-cluster algorithms are also applicable to quantum spin
models in an arbitrary magnetic field for which a similar type of sign problem 
arises. Similarly, one can solve the sign problem resulting from a complex 
action in the 2-d $O(3)$ model at non-zero chemical potential or at non-zero
vacuum angle $\theta$. The next challenge is to find applications of this 
method to QCD at non-zero baryon density. It seems likely that progress along 
the lines discussed here can be made in the quantum link D-theory formulation 
of the problem\cite{Cha97,Bro97}.

U.-J. W. likes to thank the physics department of Duke University where part 
of this work was done for its hospitality and the A. P. Sloan foundation for 
its support. This work is supported in part by funds provided by the U.S.
Department of Energy (D.O.E.) under cooperative research agreements
DE-FC02-94ER40818 and DE-FG02-96ER40945.


\begin{thebibliography}{20}
\bibitem{Evertz}
The fact that an improved estimator alone cannot solve the sign problem was 
pointed out to one of the authors by H. G. Evertz a long time ago.
\end{thebibliography}
\end{document}